\documentclass[twocolumn,aps,pra,superscriptaddress,groupedaddress,nofootinbib]{revtex4-1}
\usepackage{graphicx}% Include figure files
\usepackage{dcolumn}% Align table columns on decimal point
\usepackage{bm}% bold math
\usepackage[utf8]{inputenc}
\usepackage{amssymb,amsmath,amsthm,amsbsy,mathptmx}
\DeclareMathAlphabet{\mathcal}{OMS}{cmsy}{m}{n}
\usepackage{tabularx}
\usepackage{xcolor}
\usepackage{mathtools}
\usepackage{oubraces}
\usepackage{dsfont}
\usepackage{graphicx}
\usepackage{rotating}
\usepackage[most]{tcolorbox}
\usepackage{enumerate}
\usepackage{subfloat}
\usepackage{cleveref}

\usepackage{hyperref}% add hypertext capabilities
\def\bra#1{\mathinner{\langle{#1}|}}
\def\ket#1{\mathinner{|{#1}\rangle}}

\begin{document}

\title{Deterministic transformations of three-qubit entangled pure states}
\author{G\"{o}khan Torun}
\email{torung@itu.edu.tr}
\author{Ali Yildiz}
\email{yildizali2@itu.edu.tr}
\affiliation{Department of Physics, \.{I}stanbul Technical University, Maslak 34469, \.{I}stanbul, Turkey}
\date{\today}

%%%%
%***=====================================================ABSTRACT================================================***%
%%%%

\begin{abstract}
The states of three-qubit systems split into two inequivalent types of genuine tripartite entanglement, namely the Greenberger-Horne-Zeilinger (GHZ) type and the $W$ type. A state belonging to one of these classes can be stochastically transformed only into a state within the same class by local operations and classical communications. We provide local quantum operations, consisting of the most general two-outcome measurement operators, for the deterministic transformations of three-qubit pure states in which the initial and the target states are in the same class. We explore these transformations, originally having standard GHZ and standard $W$ states, under the local measurement operations carried out by a single party and $p$ ($p=2,3$) parties (successively). We find a notable result that the standard GHZ state cannot be deterministically transformed to a GHZ-type state in which all its bipartite entanglements are nonzero, i.e., a transformation can be achieved with unit probability when the target state has at least one vanishing bipartite concurrence.
\end{abstract}

\maketitle

%%%%
%***==============================================INTRODUCTION====================================================***%
%%%%

\section{Introduction}\label{introduction}

Quantum entanglement, as a bizarre nonclassical correlation, took its first steps into quantum theory in the works of Einstein {\emph{et al}}.~\cite{Einstein-EPR} and Schrödinger~\cite{Schrodinger}. Since then entanglement theory has flourished along with quantum information theory~\cite{Amico-Entanglement,Horodecki-QEntanglement}, and it has its roots in a wide range of exceptional discoveries such as quantum teleportation~\cite{Bennett-Teleportation}, dense coding~\cite{Bennett-densecoding}, quantum cryptography~\cite{Ekert-Cryptography}, and remote state preparation~\cite{Pati-RemoteSP}. Furthermore, over the past two decades, various studies have been presented to characterize different types of nonclassical correlations, and entanglement is still the most remarkable among all these. This assessment is due to the high performance that entanglement shows when used as a resource.

The use of entanglement as a resource~\cite{Amico-Entanglement,Horodecki-QEntanglement} in quantum information and quantum computation requires its quantification, characterization, and manipulation. Currently, bipartite entanglement is pretty well understood from all these aspects indeed, however, many problems are waiting to be solved for multipartite entanglement.
Furthermore, since multipartite entanglement consists of resource states, it outperforms bipartite entanglement~\cite{Epping-Multi,Mate-Multi,Yamasaki-Multi} in a number of different scenarios.
In this respect, entanglement manipulation of multipartite pure states by local operations and classical communications (LOCC)--free operations in the context of the resource theory of entanglement~\cite{Plenio-EntResource,Chitambar-QRTs}--is frequently seen as one of the fundamental tasks that has been widely studied in the theory of quantum information.

Traditionally, the deterministic transformation of quantum states is very important: If the initial state $\ket{\psi}$ can be deterministically transformed to a state $\ket{\phi}$, then any information task that can be done by $\ket{\phi}$ can also be done by $\ket{\psi}$. In particular, deterministic transformations can also be used to define an ordering on quantum states. Nielsen~\cite{Nielsen-Condition} used the theory of majorization and found the necessary and sufficient conditions for deterministic transformations of bipartite pure states, and some explicit protocols for these transformations were introduced in Ref.~\cite{Torun-DetBipartite}.
%For 3-partite states (refler sadece qubitse qubits yaz partite yerine), there have been studies on entanglement manipulation of %$3$-partite  states~\cite{Acin-DistillGHZ,Todd-DistillGHZ,Yildiz-OpdisWstates,Cui-Bounds,Karpat-OptFlip,Torun-canonical}
%~\cite{Vicente-EntManLOCC, Spee-PureStateTr}, though much less is known about the complete solution of the problem for $n\geq 3$.
In the multipartite case, however, the problem of the manipulation of entangled states becomes much more complicated because there are many different classes of states, even for qubits,
which cannot be converted into each other by stochastic local operations and classical communications (SLOCC)~\cite{Dur-two-inequivalent-ways,Acin-mixed-three-qubit,Moor-FourQubits}.
In the three-qubit case, the two different SLOCC class states are Greenberger-Horne-Zeilinger (GHZ) class and $W$ class states.
A general treatment for deterministic transformations of these states can be found in Refs.~\cite{Tajima-det,Turgut-Rank2Tr,Turgut-W-type}, and optimal probabilistic transformations were presented in Refs.~\cite{Acin-DistillGHZ,Todd-DistillGHZ,Yildiz-OpdisWstates,Cui-Bounds,Karpat-OptFlip,Torun-canonical}.

The notion of a maximally entangled set (MES)~\cite{Vicente-MaxSet}, which is useful for insight into quantum state transformations, is the minimal set of states such that any other state can be deterministically obtained from states in the MES by LOCC. The MES can fairly be considered as the most useful set of states for applications.  It was also shown in Ref.~\cite{Vicente-MaxSet} that
the MES is of measure zero for three-qubit states and deterministic transformations are almost never possible among fully entangled four-partite states, i.e., the MES is of full measure for four-qubit states~\cite{Spee-4Qubits}.
In general, LOCC transformations are not the largest set of operations which do not create entanglement. Hence it is reasonable to define a mathematically more manageable larger class of operations which do not create entanglement. An example is the transformation by separable operations~\cite{Vlad-SEP,Gour-SEP} which constitute a necessary but not sufficient condition for the existence of LOCC transformations. It was shown~\cite{Haibin-SepLOCC} that if the initial state $\ket{\psi}$ and final state $\ket{\phi}$ are three-qubit pure states in the GHZ class then $\ket{\psi}$ can be transformed to $\ket{\phi}$ by separable operations if and only if $\ket{\psi}$ can be transformed to $\ket{\phi}$ by deterministic LOCC. For tripartite qutrit states, however,
Hebenstreit {\it et al}.~\cite{Hebenstreit-MaxEntSet} presented an example of state transformation which is possible by separable operations but not by LOCC. Furthermore, researchers~\cite{Vicente-ResourceTheories} introduced a general resource theory of multipartite entanglement with two possible free operations: full-separability-preserving (FSP) operations and biseparability-preserving (BSP) operations. They showed that, although it is possible to obtain a transformation between different SLOCC classes by FSP operations for three-qubit states, there is no MES from which every state can be deterministically obtained by FSP operations. It was also shown that the generalized GHZ state is the MES under the BSP transformations of genuinely multipartite entangled states. Recent studies showed that, apart from the three-qubit case, nontrivial deterministic LOCC transformations among generic, fully entangled, pure states are almost never possible~\cite{Gour-AlmostMulti,Wallach-AlmostMulti}, i.e., almost no multipartite states can be either reached or deterministically converted into any other.

Even though extensive studies have been done on the LOCC transformations of multipartite states, the number of explicit protocols for these transformations is still very limited due to the large numbers of parameters in the measurement operators, even for the simplest multipartite case, i.e., three qubits.
To overcome this difficulty we use the equivalence classes of operators which were presented and used in~\cite{Torun-canonical}.
Providing a simple and practical protocol for the deterministic transformations of three-qubit entangled pure states is the subject of this paper.

In this paper we first introduce an explicit and comprehensive protocol for the deterministic LOCC transformations of a GHZ-type state into another GHZ-type state. To assess whether and how the standard GHZ state is transformed into a GHZ-type state deterministically, we use the most general local quantum operations: canonical operators for three-qubit systems. Importantly, we reveal that all GHZ class states, except the ones where all three bipartite concurrences are nonzero, can be obtained by deterministic transformations of the standard GHZ state. After that, we present local quantum operations which allow three particles to transform a $W$-type state into another $W$-type state in three steps with unit probability. These operations again consist of the most general two-outcome measurement operators. We also apply the same protocol to the standard $W$ state to show how it is transformed into a general $W$-type state.

The rest of the paper proceeds as follows. In Sec.~\ref{Sec:three-qubit} we recall some definitions for three-qubit pure states and their entanglement parameters. We provide the local measurement operators for the deterministic transformations of a GHZ-type state into another GHZ-type state in Sec.~\ref{Sec:GHZ-type}. We then present in Sec.~\ref{Sec:W-type} the local measurement operators for the deterministic transformations of a $W$-type state into another $W$-type state. In Sec.~\ref{Sec:conclusion} we conclude our work with a summary.

%%%%
%***==========================================THREE-QUBİT PURE STATES ============================================***%
%%%%

\section{Three-qubit Pure States}\label{Sec:three-qubit}

This section contains some definitions for key terms of three-qubit pure states and their entanglement parameters
that are needed for a clear understanding of the presented work.
Let us commence with the canonical form of three-qubit pure states.
Following the approach presented in Refs.~\cite{Acin-three-qubit-gen-Schmidt,Acin-three-qubit-canonical},
%one can define the canonical form of the thee-qubit pure states. Any three-qubit state
%\begin{equation}
%\ket{\phi}=\sum_{ijk}{t_{ijk}\ket{ijk}}
%\end{equation}
%defines matrices  $T_0$ and $T_1$ by
%\begin{equation}
%\ket{\phi}=\sum_{jk}{T_{0,jk}\ket{0}\ket{jk}+T_{1,jk}\ket{1}\ket{jk}}.\nonumber
%\end{equation}
%Under the unitary transformation on the first qubit,   the matrices $T_0$ and $T_1$ transform as
%\begin{eqnarray}
%T_0'&=&u_{00}^AT_0+u_{01}^AT_1, \nonumber\\
%T_1'&=&u_{10}^AT_0+u_{11}^AT_1 \quad , \quad
%u_{tz}=\braket{t|U|z}.
%\end{eqnarray}
%It is always possible to make $\textnormal{det} T_0'=0$ and the unitary transformations on the second and third qubits diagonalize %$T_0'$ which  bring the state to
one can express the canonical form of three-qubit pure states such that
\begin{eqnarray}\begin{aligned}\label{threequbitcanonical}
\ket{\psi}=&\lambda_0\ket{000}+\lambda_1 e^{i\varphi}\ket{100}+\lambda_2\ket{101}+\lambda_3\ket{110} \\
&+\lambda_4\ket{111}, \quad \left(\lambda_i \geq 0\right),
\end{aligned}\end{eqnarray}
where the coefficients $\lambda_i$ satisfy $\sum_{i=0}^4 \lambda_i^2=1$.
%Notice that the parameters $\lambda_i$ and $\varphi$ do not %uniquely determine the equivalence class while there are two solutions for $\textnormal{det} T_0'=0$. Furthermore,
It is well known that if two arbitrary states $\ket{\psi}$ and $\ket{\phi}$ are related by local unitaries, i.e., if these two states are local unitary equivalent (LUE), then they are equal (mostly written $\ket{\psi}\sim\ket{\phi}$) from the information-theoretic point of view.
%they can be obtained with certainty from each other by means of LOCC.
When deterministic transformations of entangled pure states are investigated, the LUE forms of entangled pure states constitute one of the most crucial points of the whole process. In two-qubits case, for instance, the maximally entangled pure state, $(\ket{00}+\ket{11})/\sqrt{2}$, can be transformed into the state $a\ket{00}+b\ket{11}$ with unit probability. A two-outcome measurement, carried out by one of the parties, with the measurement operators $a\ket{0}\bra{0}+b\ket{1}\bra{1}$ and $b\ket{0}\bra{0}+a\ket{1}\bra{1}$ yields one of the states $a\ket{00}+b\ket{11}$ and $b\ket{00}+a\ket{11}$, respectively.  These two states are LUE under the unitary transformation $\ket{0}\leftrightarrow\ket{1}$ on both qubits. Thus, having a LUE form of the target state in a deterministic transformation makes the problem easier to examine.
In this sense, a LUE form of the state \eqref{threequbitcanonical} was presented in Ref.~\cite{Torun-canonical},
\begin{eqnarray}\begin{aligned}\label{threequbitcanonicalprime}
\ket{\psi'}=&\lambda_0' \ket{000}+\lambda_1' e^{i\varphi'} \ket{100}+\lambda_2'\ket{101}+\lambda_3'\ket{110} \\
&+\lambda_4'\ket{111}, \quad \left(\lambda_i' \geq 0\right),
\end{aligned}\end{eqnarray}
where the coefficients $\lambda_i'$ satisfy $\sum_{i=0}^4 \lambda_i'^2=1$, as usual. As introduced in Ref.~\cite{Torun-canonical}, local unitary equivalence of the states \eqref{threequbitcanonical} and \eqref{threequbitcanonicalprime} implies
\begin{widetext}
\begin{eqnarray}\begin{aligned}\label{lambda-prime-equivalent}
\lambda_0'=&\frac{\lambda_0}{\kappa},\quad \lambda_2'=\lambda_2{\kappa}, \quad
\lambda_3'=\lambda_3{\kappa},\quad \lambda_4'=\lambda_4{\kappa}, \quad
\kappa\equiv \sqrt{\frac{\tau+C_{BC}^2}{4(\lambda_2^2+\lambda_4^2)(\lambda_3^2+\lambda_4^2)}}, \\
\lambda_1' e^{i\varphi'}=&\frac{\lambda_1}{\kappa}\left(\frac{\lambda_4^2(\lambda_2^2+\lambda_3^2+\lambda_4^2)
-\lambda_2^2\lambda_3^2}{(\lambda_2^2+\lambda_4^2)(\lambda_3^2+\lambda_4^2)}\cos(\varphi)-i\sin(\varphi)
+\frac{\lambda_2\lambda_3\lambda_4(\lambda_0^2+\lambda_1^2-\lambda_2^2-\lambda_3^2-\lambda_4^2)}
{\lambda_1(\lambda_2^2+\lambda_4^2)(\lambda_3^2+\lambda_4^2)}\right).
\end{aligned}\end{eqnarray}
As one knows, two arbitrary three-qubit pure states are LUE if and only if their entanglement parameters are the same.
In this way, the unitary invariants--entanglement parameters--can be found to be
\begin{eqnarray}\begin{aligned}\label{five-unitary-invariants}
C_{AC}&=2\lambda_0' \lambda_2'=2\lambda_0\lambda_2, \quad
C_{AB}=2\lambda_0' \lambda_3'=2\lambda_0\lambda_3, \\
C_{BC}&=2|\lambda_2'\lambda_3'-e^{i\varphi'}\lambda_1'\lambda_4'|
=2|\lambda_2\lambda_3-e^{i\varphi}\lambda_1\lambda_4|,            \\
\tau&=4\lambda_0'^2\lambda_4'^2=4\lambda_0^2\lambda_4^2.
%\tilde{\lambda}_2\tilde{\lambda}_3-\tilde{\lambda}_1\tilde{\lambda}_4\cos(\tilde{\varphi})
%=\lambda_2\lambda_3-\lambda_1\lambda_4\cos(\varphi), \nonumber \\
%J_5&=4\lambda_0^2(|\lambda_1\lambda_4e^{i\varphi}-\lambda_2\lambda_3|^2+\lambda_2^2\lambda_3^2-\lambda_1^2\lambda_4^2)
%=\lambda_0^2 C_{BC}^2+\lambda_2^2 C_{AB}^2-\lambda_1^2\tau
\end{aligned}\end{eqnarray}
\end{widetext}
where $\tau$ is a three-tangle and $C_{AB}$ is the concurrence (bipartite entanglement) between the qubits $A$ and $B$~\cite{Coffman-concurrence}, etc. Apart from these, a phase of the entanglement was introduced in Ref.~\cite{Tajima-det} such that
\begin{eqnarray}\label{entanglement-phase}
\cos \varphi_5=\frac{\lambda_0^2 C_{BC}^2+\lambda_2^2 C_{AB}^2-\lambda_1^2\tau}{C_{AB}C_{AC}C_{BC}}, \quad \varphi_5 \in [0,\pi].
\end{eqnarray}
Here the phase $\varphi_5$ is read as the entanglement phase (EP) and
%The entanglement phase is invariant with respect to local unitary operations, because all of the parameters $J_5$ and %$C_{AB}C_{AC}C_{BC}$ are.
it becomes indefinite when $C_{AB}C_{AC}C_{BC}=0$. Thus a state whose entanglement phase $\varphi_5$ is definite has been referred to as an EP-definite state and a state whose entanglement phase $\varphi_5$ is indefinite has been referred to as an EP-indefinite state~\cite{Tajima-det}.

Essentially, two arbitrary states $\ket{\psi}$ and $\ket{\phi}$ are considered to be in the same class if there is a nonzero probability of success for the both transformations $\ket{\psi}\rightarrow\ket{\phi}$ and $\ket{\phi}\rightarrow\ket{\psi}$ through SLOCC. In the three-qubit case, there are two classes of genuine tripartite entangled states which cannot be converted into each other by SLOCC, namely, GHZ and $W$ class states \cite{Dur-two-inequivalent-ways,Acin-mixed-three-qubit}.
If the three-tangle is nonzero, then the three-qubit state is of the GHZ class.
%which is defined as
%\begin{equation}
%\ket{\psi_{\textrm{GHZ}}}=\sqrt{K}(c_{\delta}\ket{0}\ket{0}\ket{0}+s_{\delta}e^{i\varphi}\ket{\varphi_{A}}\ket{\varphi_{B}}\ket{\varphi_{C}}),
%\end{equation}
%where
%\begin{eqnarray}
%\ket{\varphi_{A}}&=&c_{\alpha}\ket{0}+s_{\alpha}\ket{1},
%\end{eqnarray}
%\begin{eqnarray}
%\ket{\varphi_{B}}&=&c_{\beta}\ket{0}+s_{\beta}\ket{1},
%\end{eqnarray}
%\begin{eqnarray}
%\ket{\varphi_{C}}&=&c_{\gamma}\ket{0}+s_{\gamma}\ket{1},
%\end{eqnarray}
%and $K=(1+2c_{\delta}s_{\delta}c_{\alpha}c_{\beta}c_{\gamma}c_{\varphi})^{-1}$ ($c_{\delta}$ and $s_{\delta}$ stand for $\cos{\delta}$ %and $\sin{\delta}$, etc.). The ranges for the five parameters are  $\delta\in(0,\pi/4], \alpha,\beta,\gamma,\in(0,\pi/2]$, and %$\varphi\in[0,2\pi)$.
However, if the three-tangle is zero and
the reduced density matrices  $\rho_A\equiv$Tr$_{BC}\ket{\psi}\bra{\psi}$, $\rho_B$, and $\rho_C$ have rank 2,
then the state $\ket{\psi}$ is a $W$-type state.
A general three-qubit $W$-type state is given by
\begin{eqnarray}\label{W-type-intro}
\ket{\psi_W}=\lambda_0\ket{000}+\lambda_1\ket{100}+\lambda_2\ket{101}+\lambda_3\ket{110},
\end{eqnarray}
where all the bipartite entanglements are nonzero and $\tau=0$, i.e., $\lambda_i>0$ for $i=0,2,3$ and $\lambda_1\geq 0$.
Additionally, the standard GHZ state is given by
\begin{eqnarray}\label{pure-GHZ}
\ket{\textnormal{GHZ}}=\frac{1}{\sqrt{2}}(\ket{000}+\ket{111}),
\end{eqnarray}
and the standard $W$ state is given by
\begin{eqnarray}\label{pure-W}
\ket{W}=\frac{1}{\sqrt{3}}(\ket{001}+\ket{010}+\ket{100}).
\end{eqnarray}
These two states are completely inaccessible to each other by means of SLOCC.

To establish an effective protocol, it is crucial to start with suitable measurement operators.
The complexity of the optimal transformation of $n$-qubit ($n\geq 3$) systems is due to local quantum operations having vast numbers of parameters. The most general local measurement operators acting on qubits are $2\times2$ complex matrices
\begin{eqnarray}\begin{aligned}
M=&y_{1}e^{i\delta_1}\ket{0}\bra{0}+y_{2}e^{i\delta_2}\ket{0}\bra{1}+y_{3}e^{i\delta_3}\ket{1}\bra{0}\\
&+y_{4}e^{i\delta_4}\ket{1}\bra{1},
\end{aligned}\end{eqnarray}
where $y_k\geq 0$ and $\delta_k \in [0,2\pi)$ for $k=1,2,3,4$. Two operators $M$ and $M'$ are in the same equivalence class ($M\equiv M'$) if they both transform states in one equivalence class to states in some other equivalence class with the same probability of success. In this context, the equivalence classes of local measurements, which allows one to write the operators with the minimal number of parameters, were defined in Ref.~\cite{Torun-canonical}. Throughout this paper, while the use of the canonical operators, i.e., the most general local measurement operators, simplifies the state transformations, the canonical forms of local measurement operators~\cite{Torun-canonical} will be used. These are given by
\begin{eqnarray}\label{cano-a}
M_{A_k}=a_{00k}\ket{0}\bra{0}+a_{10k}e^{i\theta_{a_k}}\ket{1}\bra{0}+a_{11k}\ket{1}\bra{1},
\end{eqnarray}
\begin{eqnarray}\label{cano-b}
M_{B_k}=b_{00k}\ket{0}\bra{0}+b_{01k}e^{i\theta_{b_k}}\ket{0}\bra{1}+b_{11k}\ket{1}\bra{1},
\end{eqnarray}
\begin{eqnarray}\label{cano-c}
M_{C_k}=c_{00k}\ket{0}\bra{0}+c_{01k}e^{i\theta_{c_k}}\ket{0}\bra{1}+c_{11k}\ket{1}\bra{1},
\end{eqnarray}
for the parties $A$, $B$, and $C$, respectively. Here $\theta_{x_k}\in[0,2\pi)$ ($x=a,b,c$) and all the coefficients are real.
It is important to stress that to be able to apply a deterministic LOCC transformation to a given state, all the outputs are supposed to be LUE.
%The use of equivalence classes significantly reduces the number of parameters as the states
%that can be transformed into each other by local unitary transformations are equal from the information theoretic point of view.
We have two LUE states given in Eqs.~\eqref{threequbitcanonical} and \eqref{threequbitcanonicalprime}; therefore, it is required to focus on a general two-outcome local operation for the desired deterministic transformations.
To recap, the key ingredient for the protocol described in this
paper is determining the right threshold; we will consider a general two-outcome measurement of the form \eqref{cano-a}-\eqref{cano-c}, and these operations yield two states $\ket{\xi_1}$ and $\ket{\xi_2}$ which are LUE ($\ket{\xi_1}\sim\ket{\xi_2}$).

%%%%
%***===================================THREE-QUBİT PURE STATES TRANSFORMATİONS====================================***%
%%%%

%**=================================**%
%   GHZ-type states transformations   %
%**=================================**%

\section{Transformations of GHZ-type states}\label{Sec:GHZ-type}

We now proceed to examine the deterministic LOCC transformations of three-qubit GHZ-type states.
We will discuss this problem in Secs.~\ref{Subsec:measurements-one-party-only}, \ref{Subsec:measurement-two-parties-only}, and \ref{Subsec:measurements-on-three-parties}, each concerned with a certain final state.
More specifically, in Sec.~\ref{Subsec:measurements-one-party-only} the target state has only one nonzero bipartite entanglement and in Sec.~\ref{Subsec:measurement-two-parties-only} the target state has two nonzero bipartite entanglements. Section~\ref{Subsec:measurements-on-three-parties} addresses to the final state where all bipartite entanglements are nonzero.

%**************%
%one party only%
%**************%

\subsection{States with only one nonzero bipartite entanglement--local measurements by a single party}\label{Subsec:measurements-one-party-only}

The transformation under scrutiny is the following. We initially have the standard GHZ state given in Eq.~\eqref{pure-GHZ} and aim to obtain a GHZ-type state, which has only one nonzero bipartite entanglement, via local quantum operations. There are three GHZ-type states with only one nonzero bipartite entanglement:
\begin{eqnarray}\label{C-bc-neq0}
\lambda_{0}\ket{000}+\lambda_{1}\ket{100}+\lambda_{4}\ket{111}, \quad C_{BC}=2\lambda_1\lambda_4,
\end{eqnarray}
\begin{eqnarray}\label{C-ac-neq0}
\lambda_{0}\ket{000}+\lambda_{2}\ket{101}+\lambda_{4}\ket{111}, \quad C_{AC}=2\lambda_0\lambda_2,
\end{eqnarray}
\begin{eqnarray}\label{C-ab-neq0}
\lambda_{0}\ket{000}+\lambda_{3}\ket{110}+\lambda_{4}\ket{111}, \quad C_{AB}=2\lambda_0\lambda_3.
\end{eqnarray}
%As we will show, the deterministic transformation of the state \eqref{pure-GHZ} into the state \eqref{C-bc-neq0} depends solely on the %party $A$, i.e., only the local operation performed by the party $A$ will transform the state $(\ket{000}+\ket{111})/\sqrt{2}$ into %the state $\lambda_{0}\ket{000}+\lambda_{1}\ket{100}+\lambda_{4}\ket{111}$ with unit probability. The parties $B$ and $C$, likewise, %can execute the same procedure for the final states \eqref{C-ac-neq0} and \eqref{C-ab-neq0}, respectively. Thus,
The local operations carried out by a single party suffice to achieve the desired transformations.
Suppose that the party $q$ performs a local operation, consisting of a set of measurement operators $\{M_{q_k}\}$, to the GHZ state given in Eq.~\eqref{pure-GHZ}. Then the output states are obtained such that
\begin{eqnarray}\label{one-party-gen-measurement}
\ket{\psi_k}=\frac{M_{q_k}\ket{\textnormal{GHZ}}}{\sqrt{p_k}}, \quad q=A,B,C,
\end{eqnarray}
where $p_k=\bra{\textnormal{GHZ}}M_{q_k}^{\dag}M_{q_k}\ket{\textnormal{GHZ}}$. The measurement operators satisfy the normalization relation $\sum_{k}M_{q_k}^{\dag}M_{q_k}=I$, where $I$ denotes the identity operator. It should be noted that in Eq.~\eqref{one-party-gen-measurement}, while the party $q$ carries out a local measurement, the other two parties do not perform any measurement on their respective systems, e.g., for $q=B$ Eq.~\eqref{one-party-gen-measurement} should be read as $(I \otimes M_{B_k} \otimes I)\ket{\textnormal{GHZ}}/\sqrt{p_k}$.
%, i.e.,
%\begin{eqnarray}
%M_{B_k} \equiv I \otimes M_{B_k} \otimes I.
%\end{eqnarray}
In the following we will present the set of measurement operators $\{M_{q_k}\}$ for parties $A$, $B$, and $C$ successively.

%%%%%%%%%%%%%%%%%%%%%%
%First, C-bc not zero%
%%%%%%%%%%%%%%%%%%%%%%

First, consider a general two-outcome local operation on the first qubit of the state given in Eq.~\eqref{pure-GHZ} with the measurement operators given by
\begin{eqnarray}\label{pure-GHZ-MA01}
M_{A_1}=\lambda_0\ket{0}\bra{0}+\lambda_1\ket{1}\bra{0}+\lambda_4\ket{1}\bra{1},
\end{eqnarray}
\begin{eqnarray}\label{pure-GHZ-MA02}
M_{A_2}=\lambda_0'\ket{0}\bra{0}-\lambda_1'\ket{1}\bra{0}+\lambda_4'\ket{1}\bra{1},
\end{eqnarray}
where $\lambda_0'=\lambda_0/\kappa$, $\lambda_1'=\lambda_1/\kappa$, $\lambda_4'=\kappa\lambda_4$, $\kappa=\sqrt{\lambda_0^2+\lambda_1^2}/\lambda_4$, and $\sum_{k=1}^{2}M_{A_k}^{\dag}M_{A_k}=I$. The state after the measurements performed by party $A$, i.e., $\ket{\psi_k}=(M_{A_k} \otimes I \otimes I)\ket{\textnormal{GHZ}}/\sqrt{p_k}$, will be one of the states
\begin{eqnarray}\label{MA-output1}
\ket{\psi_1}=\lambda_{0}\ket{000}+\lambda_{1}\ket{100}+\lambda_{4}\ket{111},
\end{eqnarray}
\begin{eqnarray}\label{MA-output2}
\ket{\psi_2}=\lambda_0'\ket{000}-\lambda_1'\ket{100}+\lambda_4'\ket{111},
\end{eqnarray}
with probabilities $p_k=\bra{\textnormal{GHZ}}M_{A_k}^{\dag}M_{A_k}\ket{\textnormal{GHZ}}=1/2$ for $k=1,2$.
The states given in Eqs.~\eqref{MA-output1} and \eqref{MA-output2} are LUE.
The local unitary transformations
\begin{eqnarray}\label{Unitary-measurement-party-A}
U_A=\frac{-\lambda_1 I - i\lambda_0 \sigma_y}{\sqrt{\lambda_0^2+\lambda_1^2}}, \quad
U_B=i\sigma_y, \quad U_C=-\sigma_x,
\end{eqnarray}
on the qubits $A$, $B$, and $C$, respectively, will transform the state $\ket{\psi_2}$ into the state $\ket{\psi_1}$:
$(U_A\otimes U_B\otimes U_C)\ket{\psi_2}=\ket{\psi_1}$. Here $\sigma_x=\ket{0}\bra{1}+\ket{1}\bra{0}$, $\sigma_y=-i\ket{0}\bra{1}+i\ket{1}\bra{0}$, and $\sigma_z=\ket{0}\bra{0}-\ket{1}\bra{1}$ are the Pauli matrices.
While all bipartite entanglements are initially zero, by the measurement on the first qubit, the bipartite entanglement between the second and third qubits becomes nonzero ($C_{BC}=2\lambda_1\lambda_4$). However, the bipartite entanglements between the first qubit and the other two qubits is still zero ($C_{AB}$=$C_{AC}$=$0$). In other words, the measurement carried out by party $A$ has no effect on the bipartite entanglements between the particles $A-B$ and $A-C$.

%%%%%%%%%%%%%%%%%%%%%%%
%Second, C-ac not zero%
%%%%%%%%%%%%%%%%%%%%%%%

Second, consider a general two-outcome local operation on the second qubit of the state given in Eq.~\eqref{pure-GHZ}
with the measurement operators given by
\begin{eqnarray}\label{pure-GHZ-MB01}
M_{B_1}=\lambda_0\ket{0}\bra{0}+\lambda_2\ket{0}\bra{1}+\lambda_4\ket{1}\bra{1},
\end{eqnarray}
\begin{eqnarray}\label{pure-GHZ-MB02}
M_{B_2}=\lambda_0'\ket{0}\bra{0}-\lambda_2'\ket{0}\bra{1}+\lambda_4'\ket{1}\bra{1},
\end{eqnarray}
where $\lambda_0'=\lambda_0/\kappa$, $\lambda_2'=\kappa\lambda_2$, $\lambda_4'=\kappa\lambda_4$, $\kappa=\lambda_0/\sqrt{\lambda_2^2+\lambda_4^2}$, and $\sum_{k=1}^{2}M_{B_k}^{\dag}M_{B_k}=I$. The state after the measurements performed by party $B$, i.e., $\ket{\psi_k}=(I \otimes M_{B_k} \otimes  I)\ket{\textnormal{GHZ}}/\sqrt{p_k}$, will be one of the states
\begin{eqnarray}\label{MB-output1}
\ket{\psi_1}=\lambda_{0}\ket{000}+\lambda_{2}\ket{101}+\lambda_{4}\ket{111},
\end{eqnarray}
\begin{eqnarray}\label{MB-output2}
\ket{\psi_2}=\lambda_0'\ket{000}-\lambda_2'\ket{101}+\lambda_4'\ket{111},
\end{eqnarray}
with probabilities $p_k=\bra{\textnormal{GHZ}}M_{B_k}^{\dag}M_{B_k}\ket{\textnormal{GHZ}}=1/2$ for $k=1,2$.
The states given in Eqs.~\eqref{MB-output1} and \eqref{MB-output2} are LUE. The local unitary transformations
\begin{eqnarray}\label{Unitary-measurement-party-B}
U_A=i\sigma_y, \quad U_B=\frac{\lambda_2 I-i\lambda_4\sigma_y}{\sqrt{\lambda_2^2+\lambda_4^2}}, \quad
U_C=-\sigma_x,
\end{eqnarray}
on the qubits $A$, $B$, and $C$, respectively, will transform the state \eqref{MB-output2} into the state \eqref{MB-output1}.
By the measurement on the second qubit, the bipartite entanglement between the first and the third qubits becomes nonzero ($C_{AC}=2\lambda_0\lambda_2$). However, the bipartite entanglement between the second qubit and the other two qubits is zero ($C_{AB}$=$C_{BC}$=$0$).

\begin{figure}[t]
	\centering
	\includegraphics[width=.7\columnwidth]{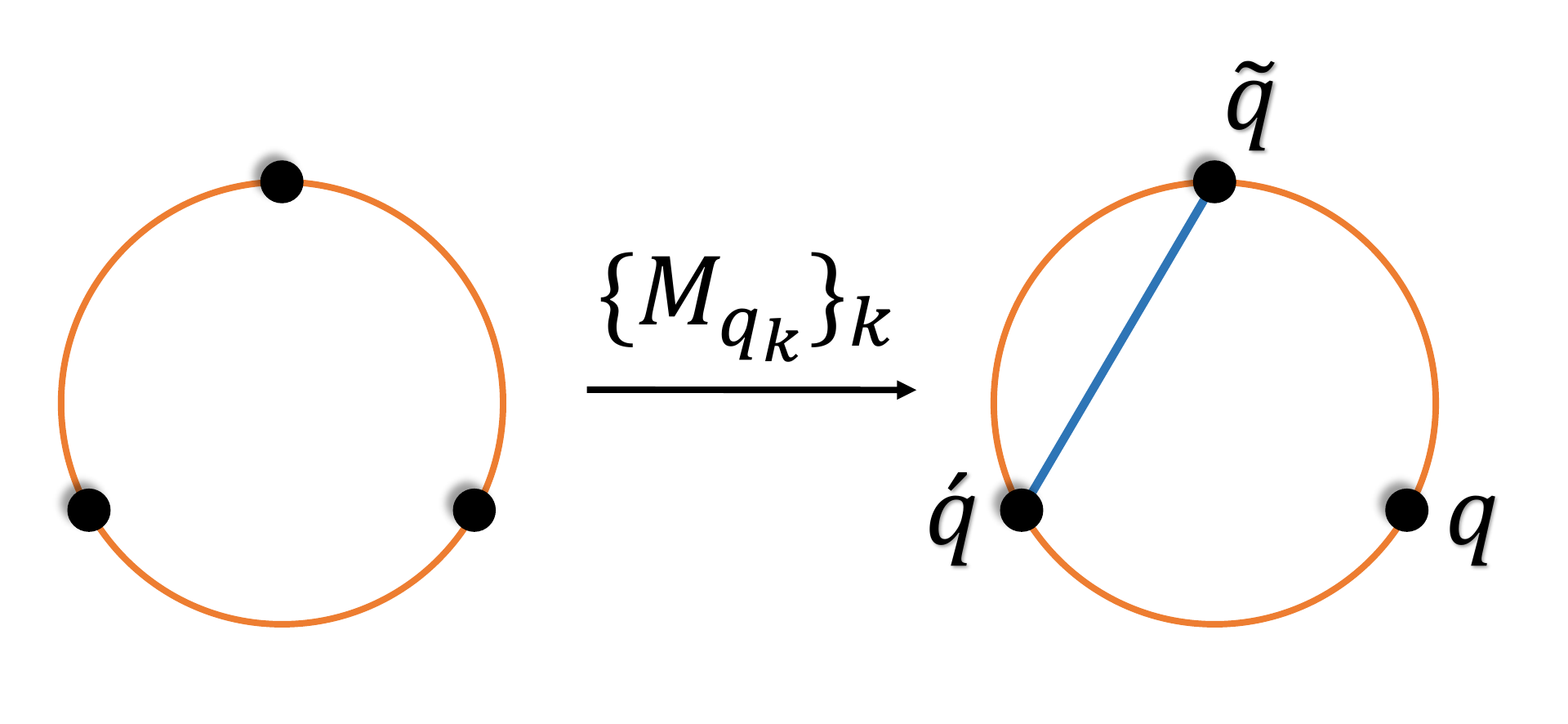}
	\caption{Pictorial representation of the deterministic LOCC transformations of tripartite GHZ-type states discussed in Sec.~\ref{Subsec:measurements-one-party-only}. Each point represents one qubit. While an orange circle connecting the three qubits denotes a three-tangle, a blue line connecting two qubits denotes bipartite entanglement \cite{Dai-ProbTeleportation}. The party $q$ performs a local operation with the measurement operators $\{M_{q_k}\}$ on the standard GHZ state. Then, for the final state, we have $C_{q\tilde{q}}=C_{q\acute{q}}=0$ and $C_{\tilde{q}\acute{q}}\neq0$.}
	\label{fig1:Cpartyqnonzero}
\end{figure}

%%%%%%%%%%%%%%%%%%%%%%
%Third, C-ab not zero%
%%%%%%%%%%%%%%%%%%%%%%

Third, consider a general two-outcome local operation on the third qubit of the state given in Eq.~\eqref{pure-GHZ}
with the measurement operators given by
\begin{eqnarray}\label{pure-GHZ-MC01}
M_{C_1}=\lambda_0\ket{0}\bra{0}+\lambda_3\ket{0}\bra{1}+\lambda_4\ket{1}\bra{1},
\end{eqnarray}
\begin{eqnarray}\label{pure-GHZ-MC02}
M_{C_2}=\lambda_0'\ket{0}\bra{0}-\lambda_3'\ket{0}\bra{1}+\lambda_4'\ket{1}\bra{1},
\end{eqnarray}
where $\lambda_0'=\lambda_0/\kappa$, $\lambda_3'=\kappa\lambda_3$, $\lambda_4'=\kappa\lambda_4$, $\kappa=\lambda_0/\sqrt{\lambda_3^2+\lambda_4^2}$, and $\sum_{k=1}^{2}M_{C_k}^{\dag}M_{C_k}=I$. The state after the measurements performed by party $C$, i.e., $\ket{\psi_k}=(I \otimes I \otimes M_{C_k})\ket{\textnormal{GHZ}}/\sqrt{p_k}$, will be one of the states
\begin{eqnarray}\label{MC-output1}
\ket{\psi_1}=\lambda_{0}\ket{000}+\lambda_{3}\ket{110}+\lambda_{4}\ket{111},
\end{eqnarray}
\begin{eqnarray}\label{MC-output2}
\ket{\psi_2}=\lambda_0'\ket{000}-\lambda_3'\ket{110}+\lambda_4'\ket{111},
\end{eqnarray}
with probabilities $p_k=\bra{\textnormal{GHZ}}M_{C_k}^{\dag}M_{C_k}\ket{\textnormal{GHZ}}=1/2$ for $k=1,2$.
The states given in Eqs.~\eqref{MC-output1} and \eqref{MC-output2} are LUE.
The local unitary transformations
\begin{eqnarray}\label{Unitary-measurement-party-C}
U_A=i\sigma_y, \quad U_B=-\sigma_x, \quad
U_C=\frac{\lambda_3 I-i\lambda_4\sigma_y}{\sqrt{\lambda_3^2+\lambda_4^2}},
\end{eqnarray}
on the qubits $A$, $B$, and $C$, respectively, will transform the state \eqref{MC-output2} into the state \eqref{MC-output1}.
By the measurement on the third qubit, the bipartite entanglement between the first and second qubits becomes nonzero ($C_{AB}=2\lambda_0\lambda_3$). However, the bipartite entanglement between the third qubit and the other two qubits is zero ($C_{AC}$=$C_{BC}$=$0$).

To sum up, while all bipartite entanglements are zero for the initial state \eqref{pure-GHZ}, when the party $q$ performs a local operation with the measurement operators $\{M_{q_k}\}$, the bipartite entanglement between the other two qubits, e.g., $\tilde{q}$ and $\acute{q}$, becomes nonzero, $C_{\tilde{q}\acute{q}}\neq0$. However, bipartite entanglement between the party $q$ and the other two qubits remains zero, $C_{q\tilde{q}}=C_{q\acute{q}}=0$, i.e., the measurement carried out by party $q$ has no effect on the bipartite entanglements between party $q$ and the remaining parties (see Fig.~\ref{fig1:Cpartyqnonzero}).
%This is consistent with the fact that any local operation on any particle cannot increase average entanglement between the particle %and other systems.
%Similarly, three-tangle cannot be increased by any local operation. In the case discussed above the initial three-tangle is $\tau_i$=$1$ and the final three-tangle is $\tau_f$=$4\lambda_0^2 \lambda_4^2$ $<1$ as expected.

%****************%
%two parties only%
%****************%

\subsection{States with only one vanishing bipartite entanglement--local measurements by two parties}\label{Subsec:measurement-two-parties-only}

We now aim to obtain a GHZ-type state which has only one vanishing bipartite entanglement, starting with the standard GHZ state given in Eq.~\eqref{pure-GHZ}, via local quantum operations.
%There are three GHZ-type states with only one vanishing bipartite entanglement:
%\begin{widetext}
%\begin{eqnarray}
%\ket{\psi}=\mu_{0}\ket{000}+\mu_1\ket{100}+\mu_2\ket{101}+\mu_4\ket{111}, \quad (C_{AC}\neq 0, \ \ C_{BC}\neq 0, \ \ C_{AB}=0),
%\end{eqnarray}
%\begin{eqnarray}
%\ket{\psi}=\mu_{0}\ket{000}+\mu_1\ket{100}+\mu_3\ket{110}+\mu_4\ket{111}, \quad (C_{AB}\neq 0, \ \ C_{BC}\neq 0, \ \ C_{AC}=0),
%\end{eqnarray}
%\begin{eqnarray}
%\ket{\psi}=\mu_{0}\ket{000}+\mu_1\ket{100}+\mu_2\ket{101}+\mu_3\ket{110}+\mu_4\ket{111}, \quad (C_{AB}\neq 0, \ \ C_{AC}\neq 0, \ \ %C_{BC}=0).
%\end{eqnarray}
%\end{widetext}
We will carry out the desired transformations in two steps, and for the first steps we will exploit the results obtained in Sec.~\ref{Subsec:measurements-one-party-only}.

Let us consider the case that in the first step of the entire transformation party $A$ performs a local operation on the state \eqref{pure-GHZ}. In that case, from the results obtained in Sec.~\ref{Subsec:measurements-one-party-only}, the state
\begin{eqnarray}\label{measurement-twoparties-outputA}
\ket{\psi}=\lambda_{0}\ket{000}+\lambda_1\ket{100}+\lambda_4\ket{111},
\end{eqnarray}
can be obtained deterministically by using the local measurement operators given in Eqs.~\eqref{pure-GHZ-MA01} and \eqref{pure-GHZ-MA02}.
%\begin{eqnarray}\begin{aligned}\label{measurement-twoparties-MA}
%M_{A_1}&=\lambda_0\ket{0}\bra{0}+\lambda_1\ket{1}\bra{0}+\frac{1}{\sqrt{2}}\ket{1}\bra{1}, \\ %M_{A_2}&=\lambda_0'\ket{0}\bra{0}-\lambda_1'\ket{1}\bra{0}+\frac{1}{\sqrt{2}}\ket{1}\bra{1},
%\end{aligned}\end{eqnarray}
%to the state given in Eq.~\eqref{pure-GHZ}.
Then, in the second step, if party $B$ performs the measurement operators
\begin{eqnarray}\begin{aligned}\label{measurement-twoparties-MB}
M_{B_1}&=\frac{\mu_0}{\sqrt{2}\lambda_0}\ket{0}\bra{0}+\mu_2\ket{0}\bra{1}+\mu_4\ket{1}\bra{1}, \\
M_{B_2}&=\frac{\mu_0'}{\sqrt{2}\lambda_0}\ket{0}\bra{0}-\mu_2'\ket{0}\bra{1}+\mu_4'\ket{1}\bra{1},
\end{aligned}\end{eqnarray}
on the state given in Eq.~\eqref{measurement-twoparties-outputA}, one of the states
\begin{eqnarray}\label{measurement-twoparties-outputB1}
\ket{\phi_1}=\mu_0\ket{000}+\mu_1\ket{100}+\mu_2\ket{101}+\mu_4\ket{111},
\end{eqnarray}
\begin{eqnarray}\label{measurement-twoparties-outputB2}
\ket{\phi_2}=\mu_0'\ket{000}+\mu_1'\ket{100}-\mu_2'\ket{101}+\mu_4'\ket{111},
\end{eqnarray}
is obtained with probabilities $p_1=p_2=1/2$, respectively, where $\lambda_0\mu_1=\lambda_1\mu_0$, $\mu_0'={\mu_0}/{\kappa}$,  $\mu_1'={\mu_1}/{\kappa}$, $\mu_2'=\kappa\mu_2$, $\mu_4'=\kappa\mu_4$, and $\kappa=\sqrt{\mu_0^2+\mu_1^2}/\sqrt{\mu_2^2+\mu_4^2}$. Also, the condition for the deterministic transformation, $\sum_{k=1}^{2}M_{B_k}^{\dag}M_{B_k}=I$, gives $\lambda_4=1/\sqrt{2}$.
The states given in Eqs.~\eqref{measurement-twoparties-outputB1} and \eqref{measurement-twoparties-outputB2} are LUE.
The local unitary transformations
\begin{eqnarray}
U_A=\frac{\mu_0\sigma_x-\mu_1\sigma_z}{\sqrt{\mu_0^2+\mu_1^2}}, \ \ U_B=\frac{\mu_2I-i\mu_4\sigma_y}{\sqrt{\mu_2^2+\mu_4^2}}, \ \
U_C=-i\sigma_y,\quad
\end{eqnarray}
on the qubits $A$, $B$, and $C$, respectively, will transform the state \eqref{measurement-twoparties-outputB2} into the state \eqref{measurement-twoparties-outputB1}.
As a result, deterministic transformations of the GHZ state~\eqref{pure-GHZ} into a GHZ-type state via local operations performed by party $A$ first and party $B$ second can be expressed as
\begin{eqnarray}\begin{aligned}\label{sum-AB-two-parties}
&\ket{\textnormal{GHZ}}=\frac{1}{\sqrt{2}}(\ket{000}+\ket{111}) \\
&\downarrow \left\{M_{A_k}\right\}_{k=1,2}, \\
&\ket{\psi}=\lambda_0\ket{000}+\lambda_1\ket{100}+\frac{1}{\sqrt{2}}\ket{111} \\
&\downarrow \left\{M_{B_k}\right\}_{k=1,2}, \\
&\ket{\phi}=\mu_{0}\ket{000}+\mu_1\ket{100}+\mu_2\ket{101}+\mu_4\ket{111},
\end{aligned}\end{eqnarray}
where the sets of measurement operators $\{M_{A_k}\}_{k=1,2}$ and $\{M_{B_k}\}_{k=1,2}$ are given in Eqs.~\eqref{pure-GHZ-MA01} and \eqref{pure-GHZ-MA02} and Eq.~\eqref{measurement-twoparties-MB}, respectively. We note that for the final state $\ket{\phi}$ given in Eq. \eqref{sum-AB-two-parties}, we have $C_{AB}=0$, $C_{AC}=2\mu_0\mu_2$, and $C_{BC}=2\mu_1\mu_4$, i.e., there is no bipartite entanglement between the first and second qubits, the qubits performing sequential measurements.

One can also consider party $C$ as the particle which performs a local operation in the second step instead of party $B$.
Then, in the second step, if party $C$ performs the measurement operators
\begin{eqnarray}\begin{aligned}\label{measurement-twoparties-MC}
M_{C_1}&=\frac{\mu_0}{\sqrt{2}\lambda_0}\ket{0}\bra{0}+\mu_3\ket{0}\bra{1}+\mu_4\ket{1}\bra{1}, \\
M_{C_2}&=\frac{\mu_0'}{\sqrt{2}\lambda_0}\ket{0}\bra{0}-\mu_3'\ket{0}\bra{1}+\mu_4'\ket{1}\bra{1},
\end{aligned}\end{eqnarray}
on the state given in Eq.~\eqref{measurement-twoparties-outputA}, one of the states
\begin{eqnarray}\begin{aligned}\label{measurement-twoparties-outputC1}
\ket{\phi_1}=\mu_0\ket{000}+\mu_1\ket{100}+\mu_3\ket{110}+\mu_4\ket{111},
\end{aligned}\end{eqnarray}
\begin{eqnarray}\begin{aligned}\label{measurement-twoparties-outputC2}
\ket{\phi_2}=\mu_0'\ket{000}+\mu_1'\ket{100}-\mu_3'\ket{110}+\mu_4'\ket{111},
\end{aligned}\end{eqnarray}
is obtained with probabilities $p_1=p_2=1/2$, respectively, where $\lambda_0\mu_1=\lambda_1\mu_0$, $\mu_0'={\mu_0}/{\kappa}$, $\mu_1'={\mu_1}/{\kappa}$, $\mu_3'=\kappa\mu_3$, $\mu_4'=\kappa\mu_4$, and $\kappa={\sqrt{\mu_0^2+\mu_1^2}}/{\sqrt{\mu_3^2+\mu_4^2}}$. Also, the condition for the deterministic transformation, $\sum_{k=1}^{2}M_{C_k}^{\dag}M_{C_k}=I$, gives $\lambda_4=1/\sqrt{2}$.
The states given in Eqs.~\eqref{measurement-twoparties-outputC1} and \eqref{measurement-twoparties-outputC2} are LUE.
The local unitary transformations
\begin{eqnarray}
U_{A}=\frac{\mu_0\sigma_x-\mu_1\sigma_z}{\sqrt{\mu_0^2+\mu_1^2}}, \ \
U_{B}=-i\sigma_y, \ \
U_{C}=\frac{\mu_3 I-i\mu_4\sigma_y}{\sqrt{\mu_3^2+\mu_4^2}}, \quad
\end{eqnarray}
on the qubits $A$, $B$, and $C$, respectively, will transform the state \eqref{measurement-twoparties-outputC2} into the state \eqref{measurement-twoparties-outputC1}.
As a result, deterministic transformations of the GHZ state \eqref{pure-GHZ} into a GHZ-type state via local operations performed by party $A$ first and party $C$ second can be expressed as
\begin{eqnarray}\begin{aligned}\label{sum-AC-two}
&\ket{\textnormal{GHZ}}=\frac{1}{\sqrt{2}}(\ket{000}+\ket{111}) \\
&\downarrow \left\{M_{A_k}\right\}_{k=1,2}, \\
&\ket{\psi}=\lambda_0\ket{000}+\lambda_1\ket{100}+\frac{1}{\sqrt{2}}\ket{111} \\
&\downarrow \left\{M_{C_k}\right\}_{k=1,2}, \\
&\ket{\phi}=\mu_{0}\ket{000}+\mu_1\ket{100}+\mu_3\ket{110}+\mu_4\ket{111},
\end{aligned}\end{eqnarray}
where the sets of measurement operators $\{M_{A_k}\}_{k=1,2}$ and $\{M_{C_k}\}_{k=1,2}$ are given in Eqs.~\eqref{pure-GHZ-MA01} and \eqref{pure-GHZ-MA02} and Eq.~\eqref{measurement-twoparties-MC}, respectively.
We note that for the final state $\ket{\phi}$ given in Eq. \eqref{sum-AC-two}, we have $C_{AC}=0$, $C_{AB}\neq0$, and $C_{BC}\neq0$, i.e., the local measurements on the first and third qubits  create bipartite entanglement between the first and second qubits and between the second and third qubits, but do not create an entanglement between the first and third qubits.

Finally, consider the case that in the first step of the entire transformation party $B$ performs a local operation on the state \eqref{pure-GHZ}. In that case, from the results obtained in Sec.~\ref{Subsec:measurements-one-party-only}, the state
\begin{eqnarray}\label{measurement-twoparties-outputB-first-step}
\ket{\psi}=\lambda_1\ket{000}+\lambda_2\ket{101}+\lambda_4\ket{111},
\end{eqnarray}
can be obtained deterministically by using the local measurement operators given in Eqs.~\eqref{pure-GHZ-MB01} and \eqref{pure-GHZ-MB02}.
%\begin{eqnarray}\begin{aligned}\label{measurement-twoparties-MB-first-step}
%M_{B_1}&=\frac{1}{\sqrt{2}}\ket{0}\bra{0}+\lambda_2\ket{1}\bra{0}+\lambda_4\ket{1}\bra{1}, \\ %M_{B_2}&=\frac{1}{\sqrt{2}}\ket{0}\bra{0}-\lambda_2'\ket{1}\bra{0}+\lambda_4'\ket{1}\bra{1},
%\end{aligned}\end{eqnarray}
%to the state given in Eq.~\eqref{pure-GHZ}.
Then, in the second step, if party $C$ performs the measurement operators
\begin{eqnarray}\begin{aligned}\label{measurement-twoparties-MBfirst-MC}
M_{C_1}&=\mu_0\ket{0}\bra{0}+\frac{\mu_1}{\sqrt{2}\lambda_2}\ket{0}\bra{1}+\frac{\mu_4}{\sqrt{2}\lambda_4}\ket{1}\bra{1}, \\
M_{C_2}&=\mu_0'\ket{0}\bra{0}-\frac{\mu_1'}{\sqrt{2}\lambda_2}\ket{0}\bra{1}+\frac{\mu_4'}{\sqrt{2}\lambda_4}\ket{1}\bra{1}
\end{aligned}\end{eqnarray}
on the state given in Eq.~\eqref{measurement-twoparties-outputB-first-step}, one of the states
\begin{eqnarray}\begin{aligned}\label{measurement-twoparties-outputC1-MBfirst}
\ket{\phi_1}=&\mu_0\ket{000}+\mu_1\ket{100}+\mu_2\ket{101}+\mu_3\ket{110}\\
&+\mu_4\ket{111},
\end{aligned}\end{eqnarray}
\begin{eqnarray}\begin{aligned}\label{measurement-twoparties-outputC2-MBfirst}
\ket{\phi_2}=&\mu_0'\ket{000}-\mu_1'\ket{100}+\mu_2'\ket{101}-\mu_3'\ket{110} \\
&+\mu_4'\ket{111},
\end{aligned}\end{eqnarray}
is obtained with probabilities $p_1=p_2=1/2$, respectively, where $\lambda_4\mu_2=\lambda_2\mu_4$, $\lambda_4\mu_1=\lambda_2\mu_3$, $\mu_1\mu_4=\mu_2\mu_3$, $\mu_0'={\mu_0}/{\kappa}$, $\mu_2'=\kappa\mu_2$, $\mu_3'=\kappa\mu_3$, $\mu_4'=\kappa\mu_4$, and $\kappa=\mu_0\mu_4/\sqrt{(\mu_2^2+\mu_4^2)(\mu_3^2+\mu_4^2)}$. Also, the condition for the deterministic transformation, $\sum_{k=1}^{2}M_{C_k}^{\dag}M_{C_k}=I$, gives $\lambda_1=1/\sqrt{2}$. The states given in Eqs.~\eqref{measurement-twoparties-outputC1-MBfirst} and \eqref{measurement-twoparties-outputC2-MBfirst} are LUE. The local unitary transformations
\begin{eqnarray}
U_{A}=-i\sigma_y,\ \
U_{B}=\frac{\mu_2\sigma_z+\mu_4\sigma_x}{\sqrt{\mu_2^2+\mu_4^2}}, \ \
U_{C}=\frac{\mu_3 I-i\mu_4\sigma_y}{\sqrt{\mu_3^2+\mu_4^2}}, \quad
\end{eqnarray}
on the qubits $A$, $B$, and $C$, respectively, will transform the state \eqref{measurement-twoparties-outputC2} into the state \eqref{measurement-twoparties-outputC1}.
As a result, deterministic transformations of the pure GHZ state into a GHZ-type state via local operations performed by party $B$ first and party $C$ second can be expressed as
\begin{eqnarray}\begin{aligned}\label{sum-BC-two}
&\ket{\textnormal{GHZ}}=\frac{1}{\sqrt{2}}(\ket{000}+\ket{111}) \\
&\downarrow \left\{M_{B_k}\right\}_{k=1,2} \\
&\ket{\psi}=\frac{1}{\sqrt{2}}\ket{000}+\lambda_2\ket{101}+\lambda_4\ket{111} \\
&\downarrow \left\{M_{C_k}\right\}_{k=1,2}, \\
&\ket{\phi}=\mu_{0}\ket{000}+\mu_1\ket{100}+\mu_2\ket{101}+\mu_3\ket{110}+\mu_4\ket{111},
\end{aligned}\end{eqnarray}
where the sets of the measurement operators $\{M_{B_k}\}_{k=1,2}$ and $\{M_{C_k}\}_{k=1,2}$ are given in Eqs.~\eqref{pure-GHZ-MB01} and \eqref{pure-GHZ-MB02} and Eq.~\eqref{measurement-twoparties-MBfirst-MC}, respectively.
We note that for the final state $\ket{\phi}$ given in Eq. \eqref{sum-BC-two}, we have $C_{BC}=0$ ($\mu_1\mu_4=\mu_2\mu_3$), $C_{AB}\neq0$, and $C_{AC}\neq0$, i.e., there is no bipartite entanglement between the second and the third qubits, the qubits performing sequential measurements.
%i.e., the local measurements on the second and the third qubits create bipartite entanglements between the first and the second %qubits, between the first and the third qubits, but do not create an entanglement between the second and the third qubits.

\begin{figure}[t]
	\centering
	\includegraphics[width=1\columnwidth]{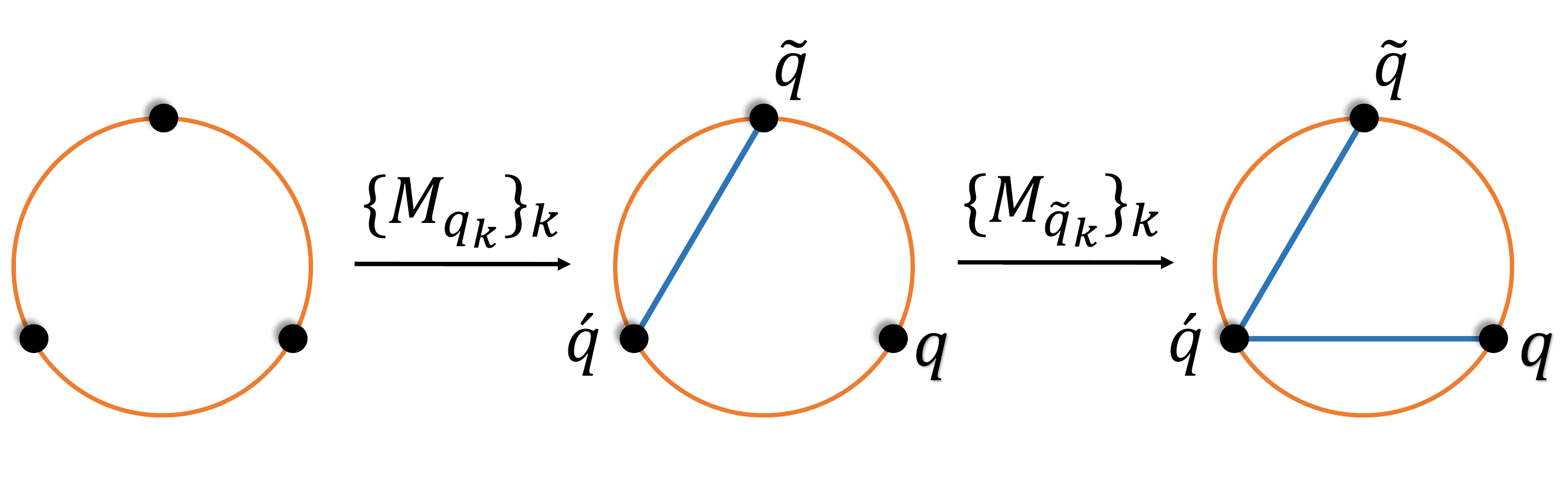}
	\caption{First, the party $q$ performs a local operation with the measurement operators $\{M_{q_k}\}$ on the standard GHZ state. Second, the party $\tilde{q}$ performs a local operation with the measurement operators $\{M_{{\tilde{q}}_k}\}$ to the output state. We finally have $C_{q\tilde{q}=0}$, $C_{q\acute{q}}\neq0$, and $C_{\tilde{q}\acute{q}}\neq0$.}
	\label{fig2:twopartyqqtilde}
\end{figure}

To sum up, when first party $q$ performs a local operation with the measurement operators $\{M_{q_k}\}$ on the state \eqref{pure-GHZ} and second party $\tilde{q}$ performs a local operation with the measurement operators $\{M_{{\tilde{q}}_k}\}$ on the state obtained from the local operation performed by party $q$ in the first step, then the bipartite entanglements between parties $q$ and $\acute{q}$ and between parties $\tilde{q}$ and $\acute{q}$ become nonzero, $C_{q\acute{q}}\neq0$ and $C_{\tilde{q}\acute{q}}\neq0$. On the other hand, the bipartite entanglement between the parties $q$ and $\tilde{q}$ is zero, $C_{q\tilde{q}=0}$ (see Fig.~\ref{fig2:twopartyqqtilde}).

%*************%
%three parties%
%*************%

\subsection{State with nonzero bipartite entanglements--local measurements by three parties}\label{Subsec:measurements-on-three-parties}

\begin{figure}[b]
	\centering
	\includegraphics[width=0.35\columnwidth]{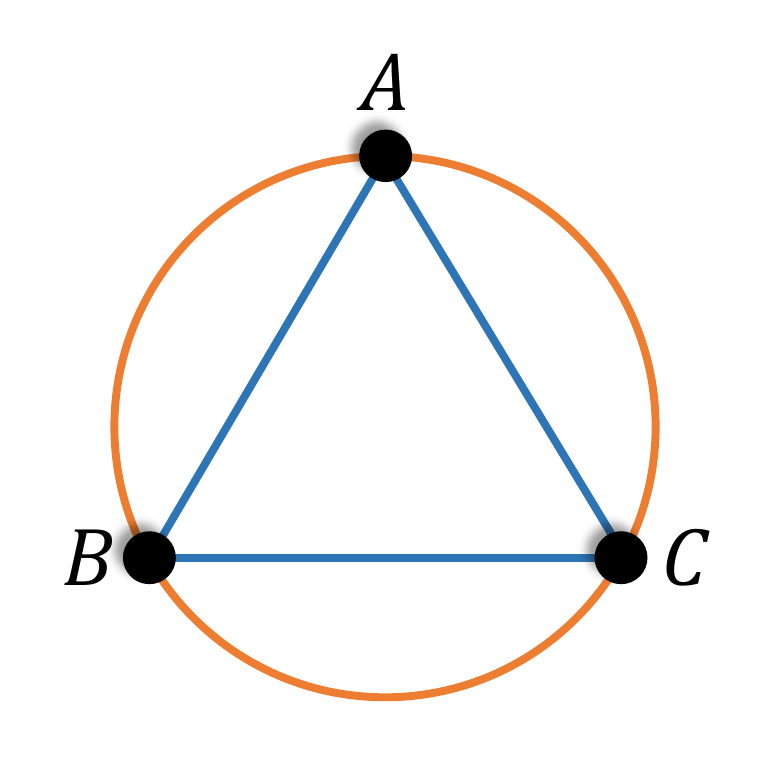}
	\caption{Pictorial representation \cite{Dai-ProbTeleportation} of the most general GHZ-type state with $\tau\neq0$, $C_{AB}\neq0$, $C_{AC}\neq0$, and $C_{BC}\neq0$.}
	\label{fig3:GHZ-nonzerobipartite}
\end{figure}

In Secs.~\ref{Subsec:measurements-one-party-only} and \ref{Subsec:measurement-two-parties-only} we have discussed the deterministic transformation of the GHZ state \eqref{pure-GHZ} (an EP-indefinite state) into an EP-indefinite GHZ-type state by LOCC. This means that the final states have at least one vanishing bipartite entanglement. We now aim to obtain the most general GHZ-type state (see Fig.~\eqref{fig3:GHZ-nonzerobipartite})--all bipartite entanglements are nonzero (an EP-definite state)--starting with the standard GHZ state given in Eq.~\eqref{pure-GHZ}, via local quantum operations.
We will try to carry out the desired transformation in three steps, and for the first two steps we will exploit the results obtained in Sec.~\ref{Subsec:measurement-two-parties-only}.
As discussed in the preceding section, the state
\begin{eqnarray}\begin{aligned}\label{three-party-outputAB}
\ket{\phi}=&\mu_{0}\ket{000}+\mu_1\ket{100}+\mu_2\ket{101}+\mu_4\ket{111},
\end{aligned}\end{eqnarray}
can be deterministically obtained by local operations performed separately by the first and second parties.
We have $\tau=4\mu_0^2\mu_4^2$, $C_{BC}=2\mu_1\mu_4$, $C_{AB}=0$, and $C_{AC}=2\mu_0\mu_2$ for the initial state \eqref{three-party-outputAB}.
Then, if party $C$ performs a two-outcome local operation by the measurement operators
\begin{eqnarray}
M_{C_k}=c_{00k}\ket{0}\bra{0}+c_{01k}e^{i\theta_{c_k}}\ket{0}\bra{1}+c_{11k}\ket{1}\bra{1}
\end{eqnarray}
%\begin{eqnarray}
%M_{C_2}=c_{002}\ket{0}\bra{0}+c_{012}e^{i\theta_{c_2}}\ket{0}\bra{1}+c_{112}\ket{1}\bra{1},
%\end{eqnarray}
for $k=1,2$ on the third qubit of the state $\ket{\phi}$ given in Eq. \eqref{three-party-outputAB}, one can obtain
\begin{widetext}
\begin{eqnarray}\begin{aligned}
(I \otimes I \otimes M_{C_k})\ket{\phi}=&c_{00k}\mu_0\ket{000}+\big[c_{00k}\mu_1+c_{01k}e^{i\theta_{c_k}}\mu_2\big]\ket{100}
+c_{01k}e^{i\theta_{c_k}}\mu_4\ket{110}+c_{11k}\mu_2\ket{101}+c_{11k}\mu_4\ket{111} \\
=& \sqrt{p_k}\ket{\zeta_k},
\end{aligned}\end{eqnarray}
where
\begin{eqnarray}\begin{aligned}\label{three-parties-output1}
\ket{\zeta_1}=\alpha_0\ket{000}+\alpha_1e^{i\alpha}\ket{100}+\alpha_2\ket{101}+\alpha_3\ket{110}+\alpha_4\ket{111},
\end{aligned}\end{eqnarray}
\begin{eqnarray}\begin{aligned}\label{three-parties-output2}
\ket{\zeta_2}=\alpha_0'\ket{000}+\alpha_1'e^{i\alpha'}\ket{100}+\alpha_2'\ket{101}+\alpha_3'\ket{110}+\alpha_4'\ket{111},
\end{aligned}\end{eqnarray}
\end{widetext}
$p_k=\bra{\phi}I \otimes I \otimes (M_{C_k}^{\dag}M_{C_k}) \ket{\phi}$ for $k=1,2$, $\theta_{c_1}=0$, $\theta_{c_2}=\pi$, $\alpha=0$, and $\alpha'=\pi$. The states given in Eqs.~\eqref{three-parties-output1} and \eqref{three-parties-output2} are LUE [for the relations between the coefficients $\alpha_i$ and $\alpha_i'$ replace $\lambda$ with $\alpha$ in Eq.~\eqref{lambda-prime-equivalent}].
Also, for the case of deterministic transformation, it is required that $\sum_{k=1}^{2}{M_{C_k}}^{\dag}M_{C_k}=I$.
The constraints on the deterministic transformation, LU equivalence of the output states $\ket{\zeta_1}$ and $\ket{\zeta_2}$ and $p_1+p_2=1$, yield
\begin{eqnarray}
\mu_0^2+\mu_1^2=\mu_2^2+\mu_4^2=\frac{1}{2},
\end{eqnarray}
for the initial state~\eqref{three-party-outputAB}, and
\begin{eqnarray}\label{alpha-concurrence}
\alpha_2\alpha_3\big(\alpha_2\alpha_3-\alpha_1\alpha_4\big)=0
\end{eqnarray}
for the final states. Equation \eqref{alpha-concurrence} can also be written such that
\begin{eqnarray}\label{concurrence-product-zero}
C_{AB} C_{AC} C_{BC}=0,
\end{eqnarray}
where $C_{AB}=2\alpha_0\alpha_3$, $C_{AC}=2\alpha_0\alpha_2$, $C_{BC}=2(\alpha_2\alpha_3-\alpha_1\alpha_4)$, and $\alpha_0\neq0$. The result ~\eqref{concurrence-product-zero} suggests that the final state must have at least one vanishing bipartite concurrence.
However, the final states \eqref{three-parties-output1} and \eqref{three-parties-output2} have nonzero bipartite entanglements (this is the case we study). Hence, deterministic transformation of the GHZ state \eqref{pure-GHZ} into a GHZ-type state is possible if the target state satisfies the condition given by Eq.~\eqref{concurrence-product-zero}. In other words, the standard GHZ state, which is an EP-indefinite state, can not be deterministically transformed to a state with all bipartite entanglements nonzero (an EP-definite state).

%%%%
%***=========================================================W-CLASS===========================================================***%
%%%%

%**===============================**%
% TRANSFORMATIONS OF W-TYPE STATES  %
%                                   %
%         *         *               %
%          *   *   *                %
%           * * * *                 %
%            *   *                  %
%**===============================**%

\section{Transformations of $W$-type states}\label{Sec:W-type}

\begin{figure}[t]
	\centering
	\includegraphics[width=0.35\columnwidth]{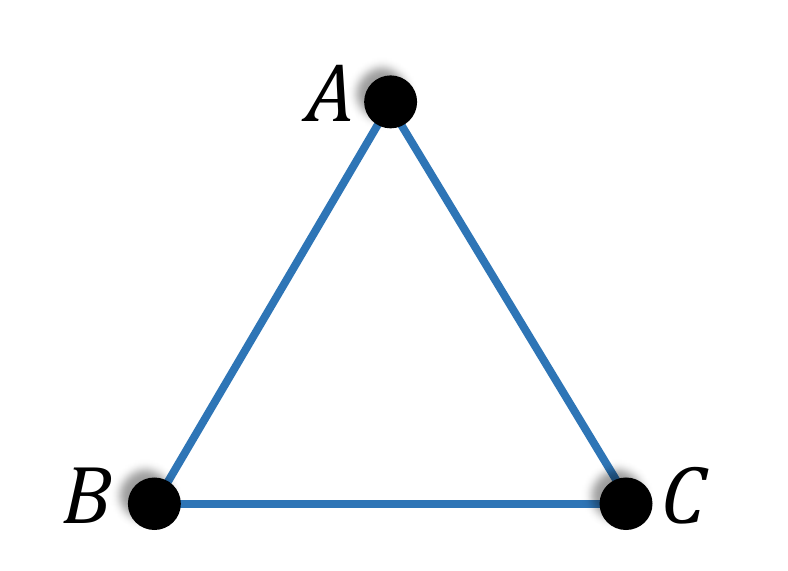}
	\caption{Pictorial representation \cite{Dai-ProbTeleportation} of the $W$-type states given in Eq.~\eqref{W-type-generic} with $\tau=0$, $C_{AB}\neq0$, $C_{AC}\neq0$, and $C_{BC}\neq0$.}
	\label{fig4:Wclass}
\end{figure}

We now provide the local measurement operators for the deterministic transformations of a $W$-type state
into another $W$-type state.
A general three-qubit $W$-type state is given by
\begin{eqnarray}\label{W-type-generic}
\ket{\psi_W}=\lambda_0\ket{000}+\lambda_1\ket{100}+\lambda_2\ket{101}+\lambda_3\ket{110},
\end{eqnarray}
where all the bipartite entanglements are nonzero and $\tau=0$, i.e., $\lambda_i>0$ for $i=0,2,3$ and $\lambda_1\geq 0$ (see Fig.~\ref{fig4:Wclass}).
As given in Ref.~\cite{Turgut-W-type}, a deterministic LOCC transformation from a $W$-type state
\begin{eqnarray}\begin{aligned}\label{wgenrl0}
\ket{\chi}=x_0 \ket{000} + x_1 \ket{100} + x_2 \ket{010} + x_3 \ket{001}
\end{aligned}\end{eqnarray}
to another $W$-type state
\begin{eqnarray}\begin{aligned}\label{wgenrl1}
\ket{\chi'}=x_0' \ket{000} + x_1' \ket{100} + x_2' \ket{010} + x_3' \ket{001}
\end{aligned}\end{eqnarray}
is possible if and only if $x_i\geq x_i'$ for $i=1,2,3$. Of course, the state \eqref{wgenrl0} can be transformed into the canonical form
\begin{eqnarray}\begin{aligned}\label{wgenrl2}
\ket{\chi}=x_1 \ket{000} + x_0 \ket{100} + x_3 \ket{101} + x_2 \ket{110},
\end{aligned}\end{eqnarray}
by the unitary transformation $\sigma_x$ on the first qubit, that is $\ket{0}_A \leftrightarrow \ket{1}_A$.
We consider the case where the initial state is $\ket{\chi}$ given in Eq.~\eqref{wgenrl2} with $x_i>0$ for $i=1,2,3$ and $x_0\geq 0$.
Then party $A$ performs a general two-outcome local operation on the state \eqref{wgenrl2}
with the canonical measurement operators
\begin{eqnarray}\begin{aligned}\label{W-type-masurement-MAfirst}
M_{A_k}=&\sqrt{p_k}\frac{\alpha_0}{x_1}\ket{0}\bra{0}
+(-1)^{k-1}\sqrt{p_{3-k}}\sqrt{1-\frac{\alpha_0^2}{x_1^2}}\ket{1}\bra{0} \quad \\
&+\sqrt{p_k}\ket{1}\bra{1},
\end{aligned}\end{eqnarray}
where $\alpha_0 \geq 0$ and $\sum_{k=1}^{2}M_{A_k}^{\dag}M_{A_k}=I$.
We then have the states
\begin{eqnarray}\begin{aligned}\label{W-type-MA-outputs}
\ket{\psi_k}=&\frac{(M_{A_k} \otimes I \otimes I) \ket{\chi}}{\sqrt{p_k}} \\
=&\alpha_{0}\ket{000}+(-1)^{k-1}\alpha_{1}\ket{100}+x_3 \ket{101}+x_2 \ket{110}
\end{aligned}\end{eqnarray}
for $k=1,2$, where we also have $x_0^2+x_1^2=\alpha_0^2+\alpha_1^2$. Here the probabilities are found to be $p_1={(\alpha_1+x_0)}/{2\alpha_1}$ and $p_2={(\alpha_1-x_0)}/{2\alpha_1}$. The local unitary transformation $\sigma_z$ on the three particles apiece allows us to transform the state \eqref{W-type-MA-outputs} for $k=2$ into the state \eqref{W-type-MA-outputs} for $k=1$, i.e., $(\sigma_z\otimes\sigma_z\otimes\sigma_z)\ket{\psi_2}=\ket{\psi_1}$.
To recap, party $A$ can transform the state \eqref{wgenrl2} into the state \eqref{W-type-MA-outputs} (for $k=1$) with unit probability via the local measurement operators \eqref{W-type-masurement-MAfirst},
\begin{eqnarray}\begin{aligned}\label{W-det-party-A}
&\ket{\chi}=x_1 \ket{000} + x_0 \ket{100} + x_3 \ket{101} + x_2 \ket{110} \\
&\downarrow \left\{\big(M_{A_k}, \: p_k=\frac{1}{2}+\frac{(-1)^{k-1}x_0}{2\alpha_1}\big)\right\}_{k=1,2}, \\
&\ket{\psi}=\alpha_{0}\ket{000}+\alpha_{1}\ket{100}+x_3 \ket{101}+x_2 \ket{110},
\end{aligned}\end{eqnarray}
%\end{empheqboxed}
where $x_1\geq\alpha_0$ and $x_0\leq\alpha_1$. The state $\ket{\psi}$ given in Eq.~\eqref{W-det-party-A} is the most general state that can be deterministically obtained by the measurements on the first qubit of the source  state $\ket{\chi}$ given in Eq.~\eqref{wgenrl2}. Next party $B$ performs a two-outcome measurement on the second qubit of the state $\ket{\psi}$ given in Eq.~\eqref{W-det-party-A} with the canonical measurement operators
\begin{eqnarray}\begin{aligned}\label{W-type-masurement-MBsecond}
M_{B_k}=&\sqrt{p_k}\ket{0}\bra{0}+(-1)^{k-1}\sqrt{p_{3-k}}\sqrt{1-\frac{\beta_3^2}{x_2^2}}\ket{0}\bra{1} \\
&+\sqrt{p_k}\frac{\beta_3}{x_2}\ket{1}\bra{1},
\end{aligned}\end{eqnarray}
where $\beta_3 \geq 0$ and $\sum_{k=1}^{2}M_{B_k}^{\dag}M_{B_k}=I$. The resulting state will then be one of the states
\begin{eqnarray}\begin{aligned}\label{W-type-MB-outputs}
\ket{\phi_k}=&\frac{(I \otimes M_{B_k} \otimes I) \ket{\psi}}{\sqrt{p_k}} \\ =&\alpha_{0}\ket{000}+{(-1)^{k-1}}\beta_{1}\ket{100}+x_3\ket{101}+\beta_3\ket{110},
\end{aligned}\end{eqnarray}
where the probabilities  are found to be $p_1={(\beta_1+\alpha_1)}/{2\beta_1}$ and $p_2={(\beta_1-\alpha_1)}/{2\beta_1}$, respectively, and $x_2^2+\alpha_1^2=\beta_1^2+\beta_3^2$. The states $\ket{\phi_1}$ and $\ket{\phi_2}$ given in Eq.~\eqref{W-type-MB-outputs} are LUE. The unitary transformation $-\sigma_z$ carried out by the party $A$, and the unitary transformation $\sigma_z$ carried out by the party $B$ will transform the state $\ket{\phi_2}$ into the state $\ket{\phi_1}$, i.e., $(-\sigma_z\otimes\sigma_z\otimes I)\ket{\phi_2}=\ket{\phi_1}$. As a result, we have
\begin{eqnarray}\begin{aligned}\label{W-det-party-B}
&\ket{\psi}=\alpha_{0}\ket{000}+\alpha_{1}\ket{100}+x_3 \ket{101}+x_2 \ket{110}  \\
&\downarrow \left\{\big(M_{B_k}, \: p_k=\frac{1}{2}+\frac{(-1)^{k-1}\alpha_1}{2\beta_1}\big)\right\}_{k=1,2}, \\
&\ket{\phi}=\alpha_{0}\ket{000}+\beta_{1}\ket{100}+x_3\ket{101}+\beta_3\ket{110},
\end{aligned}\end{eqnarray}
where $x_2\geq\beta_3$ and $\alpha_1\leq\beta_1$.
The first two qubits together can transform the state $\ket{\chi}$ given in Eq.~\eqref{wgenrl2} to the state $\ket{\phi}$ given in Eq.~\eqref{W-det-party-B} with unit probability [by combining the Eqs.~\eqref{W-det-party-A} and \eqref{W-det-party-B}].
Finally, a two-outcome measurement performed by party $C$ on the third qubit of the state $\ket{\phi}$ given in Eq.~\eqref{W-det-party-B} with the canonical measurement operators
\begin{eqnarray}\begin{aligned}\label{W-type-masurement-MCthird}
M_{C_k}=&\sqrt{p_k}\ket{0}\bra{0}+(-1)^{k-1}\sqrt{p_{3-k}}\sqrt{1-\frac{\gamma_2^2}{x_3^2}}\ket{0}\bra{1} \\
&+\sqrt{p_k}\frac{\gamma_2}{x_3}\ket{1}\bra{1},
\end{aligned}\end{eqnarray}
where $\gamma_2 \geq 0$ and $\sum_{k=1}^{2}M_{C_k}^{\dag}M_{C_k}=I$, gives one of the states
\begin{eqnarray}\begin{aligned}\label{W-type-MC-outputs}
\ket{\varphi_k}=&\frac{(I \otimes I \otimes M_{C_k})\ket{\phi}}{\sqrt{p_k}} \\
=&\alpha_{0}\ket{000}+{(-1)^{k-1}}\gamma_{1}\ket{100}+\gamma_{2}\ket{101}+\beta_{3}\ket{110},
\end{aligned}\end{eqnarray}
with probabilities $p_1={(\gamma_1+\beta_1)}/{2\gamma_1}$ and $p_2={(\gamma_1-\beta_1)}/{2\gamma_1}$, respectively. Here we have that $x_3^2+\beta_1^2=\gamma_1^2+\gamma_2^2$. The states $\ket{\varphi_1}$ and $\ket{\varphi_2}$ given in Eq.~\eqref{W-type-MC-outputs} are LUE. The unitary transformation $-\sigma_z$ carried out by the first qubit and the unitary transformation $\sigma_z$ carried out by the second qubit will transform the state $\ket{\varphi_2}$ into the state $\ket{\varphi_1}$, i.e., $(-\sigma_z\otimes\sigma_z\otimes I)\ket{\varphi_2}=\ket{\varphi_1}$. We finally have
\begin{eqnarray}\begin{aligned}\label{W-det-party-C}
&\ket{\phi}=\alpha_{0}\ket{000}+\beta_{1}\ket{100}+x_3\ket{101}+\beta_3\ket{110} \\
&\downarrow \left\{\big(M_{C_k}, \: p_k=\frac{1}{2}+\frac{(-1)^{k-1}\beta_1}{2\gamma_1}\big)\right\}_{k=1,2}, \\
&\ket{\varphi}=\alpha_{0}\ket{000}+\gamma_{1}\ket{100}+\gamma_{2}\ket{101}+\beta_{3}\ket{110},
\end{aligned}\end{eqnarray}
where $x_3\geq\gamma_2$ and $\beta_1\leq\gamma_1$.
In conclusion, we obtained the entire transformation such that
\begin{eqnarray}
\ket{\chi}\overset{\{M_{A_k}\}_k}{\longrightarrow}\ket{\psi}\overset{\{M_{B_k}\}_k}
{\longrightarrow}\ket{\phi}\overset{\{M_{C_k}\}_k}{\longrightarrow}\ket{\varphi}.
\end{eqnarray}
Here the initial state $\ket{\chi}$ given in Eq.~\eqref{wgenrl2} and the final state $\ket{\varphi}$ given in Eq.~\eqref{W-det-party-C} attest to the if and only if condition \cite{Turgut-W-type} for the deterministic transformation
$\ket{\chi}\rightarrow\ket{\varphi}$: $x_1\geq\alpha_0$, $x_2\geq\beta_3$, $x_3\geq\gamma_2$, and $x_0\leq\gamma_1$.

The canonical form of the standard $W$ state given in Eq.~\eqref{pure-W} is, by taking $x_0=0$ and $x_1=x_2=x_3=1/\sqrt{3}$ in \eqref{wgenrl2}, $(\ket{000}+\ket{101}+\ket{110})/\sqrt{3}$. Then deterministic transformation of the standard $W$ state into a $W$-type state can be written such that
\begin{eqnarray}\begin{aligned}
&\ket{W}=\frac{1}{\sqrt{3}} (\ket{000} + \ket{101} + \ket{110}) \\
&\downarrow \left\{\big(M_{A_k}, \: p_k=\frac{1}{2}\big)\right\}_{k=1,2}, \\
&\ket{\psi}=\alpha_{0}\ket{000}+\alpha_{1}\ket{100}+\frac{1}{\sqrt{3}} \ket{101}+\frac{1}{\sqrt{3}} \ket{110}  \\
&\downarrow \left\{\big(M_{B_k}, \: p_k=\frac{1}{2}+\frac{(-1)^{k-1}\alpha_1}{2\beta_1}\big)\right\}_{k=1,2}, \\
&\ket{\phi}=\alpha_{0}\ket{000}+\beta_{1}\ket{100}+\frac{1}{\sqrt{3}}\ket{101}+\beta_3\ket{110} \\
&\downarrow \left\{\big(M_{C_k}, \: p_k=\frac{1}{2}+\frac{(-1)^{k-1}\beta_1}{2\gamma_1}\big)\right\}_{k=1,2}, \\
&\ket{\varphi}=\alpha_{0}\ket{000}+\gamma_{1}\ket{100}+\gamma_{2}\ket{101}+\beta_{3}\ket{110},
\end{aligned}\end{eqnarray}
where the local measurement operators are given in Eqs.~\eqref{W-type-masurement-MAfirst}, \eqref{W-type-masurement-MBsecond}, and \eqref{W-type-masurement-MCthird}. As one can easily notice, when  party $q$ carries out a local operation the bipartite entanglement between the other two parties remains unchanged while the bipartite entanglements between party $q$ and the other two parties decrease.

%%%%
%***==============================================CONCLUSION===========================================================***%
%%%%

\section{Conclusion}\label{Sec:conclusion}

In the present paper we have set out to examine the deterministic LOCC transformations of three-qubit entangled pure states.
While an arbitrary three-qubit pure state can exist in one of the two inequivalent SLOCC classes of tripartite entanglement, we discussed the deterministic transformations in two separate sections.

We first presented local quantum operations for the deterministic transformations of a GHZ-type state into another GHZ-type state.
By using two LUE forms of three-qubit entangled pure states and the canonical forms of local measurement operators~\cite{Torun-canonical}, we introduced a simple and practical protocol, offering an alternative point of view. We originally had the standard GHZ state and applied our protocol to obtain a GHZ-type state with only one nonzero bipartite entanglement and only one vanishing bipartite entanglement. The former was achieved by a single party and the latter was achieved by the cooperation of two parties in two steps.

Next we aimed to obtain the most general GHZ-type state, the state with all bipartite entanglements nonzero.
The most significant finding to emerge from this study is that the GHZ state (and a GHZ-type state with at least one vanishing bipartite entanglement) cannot be deterministically transformed to a GHZ-type state with all bipartite entanglements nonzero. In other words, for the target state, if the bipartite entanglements satisfy the relation $C_{AB} C_{AC} C_{BC}=0$ then the deterministic transformation is possible. This result contributes to our understanding of GHZ-type state transformations.

Finally, we presented local quantum operations for the deterministic transformation of a $W$-type state into another $W$-type state.
Here we again used the canonical form of local measurement operators and achieved the transformations in three steps (i.e., with the cooperation of three parties). Each step of the entire transformation is also a deterministic transformation. Furthermore, the entire transformation gives the if and only if condition \cite{Turgut-W-type} for the deterministic transformations of $W$-type states.

%For future work: possibility to use some form of canonical operations for manipulation of mixed 3-qubit states?? sınıflandırmasına bak kaçsınıf var vs

%%%%
%***====================================================ACKNOWLEDGMENTS=================================================***%
%%%%

%\begin{acknowledgments}
%G.T. acknowledges financial support from the Scientific and Technological Research Council of Turkey (TUBITAK).
%\end{acknowledgments}

%%%%
%***==============================================REFERENCES============================================================***%
%%%%

%\bibliography{biblio}
%\bibliographystyle{apsrev}

%%%%
%***====================================================================================================================***%

\end{document}